\documentclass[aps,twocolumn,prb,showpacs,floatfix,superscriptaddress]{revtex4}

\usepackage{amsmath,fixmath}
\usepackage{graphicx}
\usepackage{acronym}
\usepackage{amssymb}
\usepackage{bm}

\newcommand{\e}{\mathrm{e}}
\newcommand{\dl}{d_\mathrm{l}}
\newcommand{\Tc}{T_\mathrm{c}}

\begin{document}
%\begin{frontmatter}
  
\title{Critical properties of the four-state Commutative Random Permutation
Glassy Potts model in three
and four dimensions}

\author{L.~A.~Fern\'andez} 
\affiliation{Departamento de F\'{\i}sica Te\'orica I, Facultad de
  F\'{\i}sicas, U. Complutense, 28040 Madrid, Spain}
\affiliation{Instituto de Biocomputaci\'on y F\'{\i}sica de Sistemas
  Complejos (BIFI) 50009 Zaragoza, Spain}
\author{A.~Maiorano}
\affiliation{Instituto de Biocomputaci\'on y F\'{\i}sica de Sistemas
  Complejos (BIFI) 50009 Zaragoza, Spain}
\affiliation{Dipartimento di Fisica, U. di Ferrara, I-44100 
Ferrara, Italy} 
\author{E.~Marinari}
\affiliation{Dipartimento Fisica, INFN and INFM,
Sapienza Universit\`a di Roma, 00185 Roma, Italy}
\author{V.~Martin-Mayor} 
\affiliation{Departamento de F\'{\i}sica Te\'orica I, Facultad de
  F\'{\i}sicas, U. Complutense, 28040 Madrid, Spain}
\affiliation{Instituto de Biocomputaci\'on y F\'{\i}sica de Sistemas
  Complejos (BIFI) 50009 Zaragoza, Spain}
\author{D.~Navarro}
\affiliation{Instituto de Investigaci\'on en Ingenier\'ia de Arag\'on 
  (I3A), U. de Zaragoza, 50018 Zaragoza, Spain}
\affiliation{Dep. de Ing. Electr\'onica y 
  Comunicaciones, C. Pol. Superior, U. de 
  Zaragoza, 50018 Zaragoza, Spain}
\author{D.~Sciretti}
\affiliation{Instituto de Biocomputaci\'on y F\'{\i}sica de Sistemas
  Complejos (BIFI) 50009 Zaragoza, Spain}
\affiliation{Departamento de F\'{\i}sica Te\'orica, Facultad de
  Ciencias, U. de Zaragoza, 50009 Zaragoza, Spain}
\author{A.~Taranc\'on}
\affiliation{Instituto de Biocomputaci\'on y F\'{\i}sica de Sistemas
  Complejos (BIFI) 50009 Zaragoza, Spain}
\affiliation{Departamento de F\'{\i}sica Te\'orica, Facultad de
  Ciencias, U. de Zaragoza, 50009 Zaragoza, Spain}
\author{J.~L.~Velasco}
\affiliation{Instituto de Biocomputaci\'on y F\'{\i}sica de Sistemas
  Complejos (BIFI) 50009 Zaragoza, Spain}
\affiliation{Departamento de F\'{\i}sica Te\'orica, Facultad de
  Ciencias, U. de Zaragoza, 50009 Zaragoza, Spain}
\date{\today}

\begin{abstract}
  We investigate the critical properties of the four-state commutative
  random permutation glassy Potts model in three and four dimensions
  by means of Monte Carlo simulation and of a finite size scaling
  analysis.  Thanks to the use of a field programmable gate array we
  have been able to thermalize a large number of samples of systems
  with large volume.  This has allowed us to observe a spin-glass
  ordered phase in $d\!=\!4$ and to study the critical properties of
  the transition. In $d\!=\!3$, our results are consistent with the
  presence of a Kosterlitz-Thouless transition, but also with
  different scenarios:
  transient effects due to a value of the lower critical dimension
  slightly below $3$ could be very important.

\end{abstract}

\maketitle

%\end{keyword}
%\end{frontmatter}

%\maketitle

%%%%%%%%%%%%%%%%%%%%%%%%%%%%%%%%%%%%%%%%%%%%%%%%%%%%%%%%%%%%%%%%%%
\section{Introduction}

In the last years, spin-glass models without spin-inversion symmetry
\cite{PSPIN,ELDER1,ELDER2,GROSS,CHIRAL-POTTS,GLASSY-POTTS,FRANZETAL,CGG,KZ}
have received a large amount of attention: probably the main reason
for this big effort is that they are thought to describe structural
glasses that in nature, as opposed to spin glasses, do not enjoy this
symmetry.  One of them, the Ferro-Potts-Glass
(FPG),\cite{ELDER1,ELDER2} is a very direct generalization of the
Ising Edwards-Anderson spin glass: the spins can take $p$ different
values, and two neighboring spins contribute to the total energy a
factor $-J_{ij}$ if they are in the same state and a factor $+J_{ij}$
if they are in different states.  The bonds $J_{ij}$ are quenched
random variables that can be distributed, for example, under a
Gaussian or under a bimodal distribution.  In the FPG, as we will
discuss better in the following, the missing spin-inversion symmetry
has the collateral effect of allowing the existence of a ferromagnetic
phase at low values of the temperature (this is why we define it
Ferro-Potts-Glass): because of this possible contamination the
analysis of the glassy critical points of the model can potentially
become very complex, and even lead to misleading conclusions. In fact,
as we will discuss below, progress can be expected from the
consideration of more refined models, where a gauge symmetry forbids
the ferromagnetic phase.

The FPG is a candidate for describing orientational glasses: a
$p$-state spin models a quadrupole moment which can be directed in $p$
(discrete) directions.\cite{BINDER-REGER} However, its main interest
is maybe originated from some of the properties of its infinite-range
version: for $p>4$, for example, the mean field FPG undergoes a glass
transition\cite{GROSS} where the order parameter is
discontinuous.\cite{PARISI} A number of different lattice
models,\cite{REM,PSPIN,ELDER1,ELDER2,CHIRAL-POTTS,GLASSY-POTTS,FRANZETAL,CGG}
in other words, can be analyzed to clarify the finite-range behavior
of systems showing the equilibrium properties typical of glasses: it
is also important to remember that a number of important connections
have been found\cite{KIRK1,KIRK2} between the mean-field dynamical
equations of the model and the mode-coupling theory of the structural
glass transition,\cite{GOTZE,BOUCHAUD} that describes the evolution of
the density correlations in a supercooled liquid above the dynamical
transition temperature.

Even if the mean-field results can be an important starting point, in
a next step, since real systems have short-range interactions, it is
important to study finite dimensional systems.  Great part of the
mainly numerical effort has been focused on the $p\!=\!3$ model in
$d\!=\!3$, to model a realistic quadrupolar glass.\cite{BINDER} The
first numerical
studies\cite{BAN1,BAN2,SCHE1,SCHE2,REUHL,CARMESIN,SINGH,LOBE} found
that the lower critical dimension $\dl$ is close to $3$.  In a
numerical study with a zero-temperature scaling approach, Banavar and
Cieplak\cite{BAN1} suggested that the FPG with Gaussian couplings has
a $\dl$ slightly greater than $3$, while the FPG with bimodal
couplings has a $\dl$ slightly below $3$ (but such a measurement had
large intrinsic errors).  A few months later Monte Carlo
simulations\cite{SCHE1,SCHE2} hinted that the transition seems to take
place at a temperature compatible with $\Tc\!=\!0$ for both families
of couplings, which suggested indeed that $\dl\!=\!3$.  Further
simulations in the bimodal\cite{REUHL} and Gaussian\cite{CARMESIN}
models were consistent with these results, although one could not
exclude the possibility of $\Tc$ being small but larger than zero.  A
later study based on a high-temperature expansion\cite{SINGH,LOBE} did
not allow to reach a final conclusion.  Only recently we start to have
clearer evidences about the situation: a large scale numerical study,
based on a finite size scaling analysis of the correlation length
\cite{YOUNG}, gives what looks like a reliable evidence of a
transition to a glass phase at finite $\Tc$, making in this way a
strong case for $\dl$ being slightly below $3$ for the three-state
FPG.

Another interesting model that has been studied in detail is the
$p\!=\!10$ model in $d\!=\!3$, because of the intrinsic interest of
the limit of a large number of states. Old\cite{BRA1,BRA2} and
recent\cite{YOUNG}numerical simulations seem to suggest that there is
no spin glass transition at finite temperature (but all the warnings
about the dangers of ferromagnetic effects at low $T$ in this model
stay in effect).  This finding is in marked contrast with the
predictions of mean field theory that indeed undergoes two
transitions:\cite{KIRK1,KIRK2} new models could be useful to
understand better the connections among the mean field and the finite
dimensional picture, and for example Potts-glass models with
medium-range interactions\cite{BRA1,BRA2} could be relevant at this
effect.

It has also been argued\cite{EASTWOOD} (although some controversy
exists\cite{BRA2}) that the choice of the coupling distribution might
be relevant in removing the phase transition on the $p\!=\!10$ model. 
The disease of the FPG that we have discussed before is
the designated culprit: the lack of the spin inversion symmetry (which
in Ising spin glasses is connected to a gauge symmetry that forbids a
spontaneous magnetization\cite{TOULOUSE}) allows ferromagnetic
ordering at low temperatures.\cite{ELDER1,ELDER2}  A partial relief to
this problem can be obtained by using a distribution of couplings
non-symmetric around zero,\cite{BRA1,YOUNG} but this choice does not
recover the lost (important) gauge invariance.

A different (and natural) definition of a frustrated Potts model
containing quenched disorder, the Random Permutation Potts Glass
(RPPG), has been introduced a few years ago.\cite{GLASSY-POTTS} The
key point of the RPPG (and of the similar model where only a set of
possible couplings is allowed, the Commutative Random Permutation
Potts Glass, CRPPG, where an additional symmetry is very useful to
help checking thermalization, see \ref{MODEL}) is that it retains the
gauge invariance which prevents Ising spin-glasses from entering
ferromagnetic ordering at low temperature.  The same
paper\cite{GLASSY-POTTS} analyzed numerically the $p\!=\!4$,
four-dimensional model (both in the RPPG and in the CRPPG versions) on
lattices of volume $V\!=\!4^4$ and $V\!=\!5^4$.  The two models were
found to exhibit the same critical behavior, with a glassy phase
characterized by a divergence of the overlap susceptibility.  A
preliminary value of $\gamma$ was estimated from that divergence, and
the critical temperature was obtained from the analysis of the Binder
parameter: the critical behavior was found to be reached under a
discontinuity, that was related to the one observed in the Random
Energy Model.\cite{REM} It is also interesting to note that
Carlucci\cite{CARLUCCI} has discussed the relation connecting the
(C)RPPG and the Chiral Potts model introduced by Nishimori and
Stephen,\cite{CHIRAL-POTTS} which in mean field shows the same type of
transition for $p>4$.\cite{GROSS,CARLUCCI} The authors of
Ref. \onlinecite{GLASSY-POTTS} also present a dynamical study of their
models, and they observe clear aging effects.

In this work we investigate, by means of Monte Carlo simulation and
Finite-Size Scaling analysis, the critical properties of the three and
of the four dimensional $p\!=\!4$ CRPPG. In $d\!=\!3$, the finite-size
behavior makes possible that the system undergoes a Kosterlitz-Thouless
transition, although a $\dl$ barely lower than $3$ is surely
compatible with the significance of our numerical data.  In $d\!=\!4$,
we confirm the existence of the spin-glass transition reported in
Ref.~\onlinecite{GLASSY-POTTS}, but the use of a field programmable
gate array (FPGA) computer (see the appendix and
Ref.~\onlinecite{IANUS2}) allows us to obtain more accurate estimates
of the critical exponents, universal dimensionless quantities and non
universal critical couplings of the model.

The remaining part of this work is organized as follows. In
Section~\ref{MODEL} we define the model and comment on its symmetries.
We describe the relevant observables in Section~\ref{OBSERVABLES}.
Section~\ref{MONTECARLOSECT} is devoted to a discussion of the
numerical methods: the details of the simulations are described in
Section~\ref{SIMULATIONS} and the Finite-Size Scaling method in
Section~\ref{FSS}. Further details about the computation 
are given in Section~\ref{COMP_DETAILS},
while the problem of thermalization is addressed in
Section~\ref{THERMALIZATION}. The results for the $d\!=\!3$ model are
discussed in Section~\ref{RESULTS3d}, and those for $d\!=\!4$ are
discussed in Section~\ref{RESULTS4d}. We present our conclusions in
Section~\ref{CONCLUSIONS}.
In 
the Appendix
we give details about the FPGA and
about how they have actually been used.

\section{The model}\label{MODELSECT}

\subsection{Model and symmetries}\label{MODEL}

We consider a system of spins $\{\sigma_i\}$
defined on a $d\!=\!3$ (and $d\!=\!4$) dimensional simple cubic lattice 
of linear size \textit{L} (volume $V=L^d$) and
periodic boundary conditions. 
The Hamiltonian is:
\begin{equation}\label{Ham}
  H\equiv - \sum_{<i,j>}\delta_{\sigma_i,\Pi_{ij}(\sigma_j)}\;,
\end{equation}
where the sum runs over all pairs of nearest neighboring sites.  The
spins can take the values $\{0,1,2,3\}$, and $\Pi_{ij}$ are quenched
permutations of $\{0,1,2,3\}$, defined on the links of the
lattice.\cite{GLASSY-POTTS,MPR} We define our quenched couplings (to
implement the commutative model of Ref. \onlinecite{GLASSY-POTTS}) by extracting
random permutations of $(0,1,2,3)$ that commute with our ``reference
permutation''
$R\!=\!(0,1,2,3)\rightarrow(2,3,0,1)$. Only links
from $i$ to $j$ such that $\sigma_i\!=\!\Pi_{ij}(\sigma_j)$ give a
non-zero contribution to the energy.  The RPPG and CRPPG are deeply
connected\cite{CARLUCCI} to the Chiral-Potts model analyzed by
Nishimori and Stephen.\cite{CHIRAL-POTTS}

The symmetry with respect to the reference permutation $R$ helps in
defining an order parameter $q$ governed by a probability
distribution symmetric under $q\to - q$ (this turns out to
be crucial for checking that the system has reached thermal
equilibrium\cite{GLASSY-POTTS}).  We define two copies of the system
(two real replicas) $\{\sigma^{(1)}_i\},\{\sigma^{(2)}_i\}$ and we
allow them to evolve independently at the same temperature and the
same realization of quenched random couplings $\Pi_{ij}$. The {\em
  modified} overlap between the two replicas at site $i$ is defined as
\begin{equation}
q_i =
\left\{
\begin{array}{rl}
1 & \mbox{if } \sigma^{(1)}_i=\sigma^{(2)}_i \;,\\
-1 & \mbox{if } \sigma^{(1)}_i\ne\sigma^{(2)}_i
\mbox{ and } 
\sigma^{(1)}_i=(\sigma^{(2)}_i+2)~\mathrm{mod}~2\;, \\
0 & \mbox{elsewhere}\;.
\end{array}
\right.
\end{equation}

%%%%%%%%%%%%%%%%%%%%%%%%%%%%%%%%%%%%%%%%%%%%%%%%%%%%%%%%%%
\subsection{Observables}\label{OBSERVABLES}

The main quantities that we will consider here are defined
in terms of the Fourier transform of $q_i$:
\begin{equation}
  \hat{q}(\vec k)=\frac{1}{V}\sum_i 
  \e^{-\mathrm{i} \vec k\cdot\vec{r}_i} q_i\;.
\end{equation}
The momentum space propagator is defined from the
relation: 
\begin{equation}
G(\vec k)=V\langle \overline{\hat{q}(\vec k)^2\rangle}\;.
\end{equation}
In the thermodynamic limit and at the critical point, the propagator is
expected to have poles at $\vec k\!=\!\vec 0$:
\begin{equation}
G(\vec k)\approx
\frac{Z\xi^{-\eta}}{(\vec k)^2+\xi^{-2}}\;,
\end{equation}
where the correlation length $\xi$ diverges at the critical point, and
$\xi\Vert\vec k\Vert\ll 1$. 
We also define the non-connected susceptibility:
\begin{equation}
\chi = G(\vec 0)\;.
\end{equation}
On a finite lattice an extremely useful definition of the correlation length
can be obtained from the discrete derivative of $G(\vec k)$. Using $\vec k\!=\!(2\pi/L) \vec e_\mu$, where $\vec e_\mu$ belongs to the canonical
Cartesian basis, one obtains:\cite{COOPER,AMIT}
\begin{equation}
\xi=\left(\frac{G(\vec 0)/G(\vec k)-1}{4\sin^2(\pi/L)}\right)^{1/2}\;.
\end{equation} 
We also compute and analyze the cumulant:
\begin{equation}
  U_4\equiv \frac{\overline{\langle\hat q(\vec
      0)^4\rangle}}{\overline{\langle\hat q(\vec
      0)^2\rangle^2}}\;.
\label{def_binder}
\end{equation}
We define the energy as:
\begin{equation}
E=\frac{4}{3 V d}\langle {H} \rangle -\frac{1}{3} \ ,
\end{equation}
so that it lays in the $[0,1]$ interval.
When we need to estimate the derivative with respect to $\beta$ of
an observable $O$, we estimate it by measuring the connected
correlation function $\langle O\; H\rangle_c$. 
Bias-corrected\cite{FSS-CHECK} reweighting
techniques\cite{FSS,BALLESTEROS,AMIT}
allow us to use the numerical data taken at temperature $T$
to compute expectation values at nearby temperature values $T'$,
and to get in this way estimates that cover all the relevant part of
the critical region.

%%%%%%%%%%%%%%%%%%%%%%%%%%%%%%%%%%%%%%%%%%%%%%%%%%%%%%%%%%%%%%%
\section{Numerical methods}\label{MONTECARLOSECT}

%%%%%%%%%%%%%%%%%%%%%%%%%%%%%%%%%%%%%%%%%%%%%%%%%%%%%%%%%%%%%%%
\subsection{Simulations}\label{SIMULATIONS}

In the $d\!=\!3$ model we have analyzed lattices of linear sizes
$L\!=\!6, 8, 10$ and $16$. The critical behavior of the model (see
Section~\ref{RESULTS3d}) has suggested to simulate a wide range of
values of $\beta$, ranging from $1.5$ to $2.7$. We have analyzed
between $200$ and $400$ different samples of the smaller systems and
around $1000$ samples for $L\!=\!16$.

In $d\!=\!4$, we have analyzed lattices of linear sizes $L\!=\!8, 12,$
and $16$, with $\beta$ ranging from $1.385$ to $1.5$.  The main
computer effort has been accomplished around $\beta\!=\!1.405$ and
$\beta\!=\!1.41$, close to the critical point. At these temperatures,
we have simulated $1000$ samples for $L\!=\!8$ and $2000$ samples for
$L\!>8$. For the other $\beta$ values we have simulated between $200$
and $400$ samples. We have also analyzed $50$ samples of the system
deep into the low-temperature region, at $\beta\!=\!1.5$.

%%%%%%%%%%%%%%%%%%%%%%%%%%%%%%%%%%%%%%%%%%%%%%%%%%%%%%%%%%%%%%%%%%%%%%%%%
\subsection{Finite size scaling}\label{FSS}

We give here a few details about the finite size scaling approach that
we have used for our analysis. When using the \textit{quotient
  method}\cite{RP2D3-LETTER,RP2D3-LONG,AMIT} one compares the mean
value of an observable $O$, in two systems of sizes $L_1$ and $L_2$,
using the value $\beta$ where the correlation length in units of the
lattice sizes coincides for both systems. If, for the infinite volume
system, $\langle O\rangle(\beta)\propto|\beta
-\beta_\mathrm{c}|^{-x_O}\,,$ the basic equation of the quotient
method is:
\begin{equation}
\begin{array}{rcl}
  Q_O^{L_1,L_2}&\equiv&\displaystyle\left.\frac{\overline{\langle
        O(\beta,L_2) \rangle}}{\overline{\langle O(\beta,L_1)
        \rangle}}\right|_{\frac{\xi(L_2,\beta)}{\xi(L_1,\beta)}=\frac{L_2}{L_1}}\\
  &=& \displaystyle\left( \frac{L_2}{L_1}
  \right)^{{x_O}/\nu}(1+A_OL_1^{-\omega}+\ldots)\;,
\label{QUOTIENTS-FORMULA}
\end{array}
\end{equation}
where the dots stand for higher-order scaling corrections, $\nu$ is
the correlation length critical exponent, $\omega$ is the (universal)
first irrelevant critical exponent, and $A_O$ is a non universal
amplitude. 

Just below the lower critical dimension, at a distance $\epsilon$, the
critical exponent $1/\nu$ is expected to be of order $\epsilon$. This
means that, for a limited range of lattice sizes, the slope of the
$\xi/L$ curves at $\Tc$ grows very slowly (almost logarithmically)
with $L$. This could make life hard for a numerical study where one
looks for a crossing of the $\xi/L$ curves, since the curves for the
different lattice sizes would be basically parallel in the critical
region. In other words, distinguishing a {\em merging} of the $\xi/L$
curves from a {\em crossing} becomes very hard. If one works precisely
at the lower critical dimension (i.e. $\epsilon\!=\!0$), one may
expect that one of two mutually excluding scenarios is realized.  If
$\Tc\!=\!0$, the curves for $\xi/L$ would not join (if plotted versus
$1/T$, the curves for lattices of size $L$ and $2L$ should displace
uniformly by a $L$-independent amount).  On the other hand, if $\Tc>0$
one would have a Kosterlitz-Thouless picture, where the curves for
$\xi/L$ merge for all $T<\Tc$. It is clear that distinguishing a
Kosterlitz-Thouless scenario from $\epsilon>0$ but very small is
numerically challenging.

The most precise way of extracting the critical point
$\beta_\mathrm{c}$ is to consider the crossing point of dimensionless
quantities such as $\xi/L$ and $U_4$. When comparing their values in
two systems of size $L_1$ and $L_2$, one finds that they take a common
value at
\begin{equation}
  \beta^{L_2,L_1}_\mathrm{c}=\beta_\mathrm{c}+ B \frac{1 -
    (L_2/L_1)^{-\omega}}{(L_2/L_1)^{1/\nu}-1} L_1^{-\omega -1/\nu} +
  \ldots\,,\label{ajuste}
\end{equation}
The non universal amplitude $B$ depends on the dimensionless
quantity that one considers.
\begin{table}[b]
% \begin{center}
\begin{ruledtabular}
\begin{tabular}{ccccccc}
 $L$ & $\beta$ & $N_\mathrm{samples}\times 10^2$ & EMCS$\times 10^6$ &
 EMCS/meas. \\
 \hline
 \hline
 6 & 1.6 & 2 & 4 & 40 \\
 \hline
 6 & 2.0 & 2 & 4 & 40 \\
 \hline
 6 & 2.4 & 4 & 4 & 40 \\
 \hline
 8 & 1.6 & 2 & 4 & 40 \\
 \hline
 8 & 1.8 & 2 & 8 & 40 \\
 \hline
 8 & 2.0 & 4 & 8 & 40 \\
 \hline
 8 & 2.4 & 4 & 4 & 40 \\
 \hline
 10 & 1.5 & 2 & 4 & 40 \\
 \hline
 10 & 1.8 & 2 & 12 & 40 \\
 \hline
 10 & 2.0 & 2 & 12 & 40 \\
 \hline
 10 & 2.2 & 4 & 12 & 40 \\
 \hline
 10 & 2.4 & 4 & 24 & 40 \\
 \hline
 16 & 1.8 & 10 & 60 & $5\times 10^5$ \\
 \hline
 16 & 2.0 & 10 & 60 & $5\times 10^5$ \\
 \hline
 16 & 2.2 & 10 & 60 & $5\times 10^5$ \\
 \hline
 16 & 2.4 &  9 & 600 & $2\times 10^6$ \\
\end{tabular}
 %\end{center}
\end{ruledtabular}
\caption{For each lattice size of the $d\!=\!3$ model, we show the
  simulated temperatures, number of samples, number of EMCS per sample
  and EMCS per measurement.}\label{tab:sim_3d}
\end{table}

%%%%%%%%%%%%%%%%%%%%%%%%%%%%%%%%%%%%%%%%%%%%%%%%%%%%%%%%%%%%%
\subsection{Computational details}\label{COMP_DETAILS}

In order to compute equilibrium expectation values we update the spins
with a sequential Metropolis algorithm, we bring them to equilibrium
and during the equilibrium dynamics we measure the interesting
physical quantities.  Thanks to our optimized FPGA based processor we
have been able to run large scale simulations: for example thanks to
strong thermalization tests we can be sure that we have thermalized systems
of volume $V=16^3$ and $V=16^4$ at high $\beta$ values, already deep
in the broken phase.  We define an \textit{elementary Monte Carlo
sweep} (EMCS) as $V$ sequential trial updates of lattice spin
(considered in lexicographic order).  To produce the needed
pseudo-random numbers we use the Parisi-Rapuano shift register
method.\cite{PARISI-RAPUANO}

The $d\!=\!3$ small lattices, from $L\!=\!6$ to $10$, have been simulated at
the cluster of the Instituto de Biocomputaci\'on y F\'{\i}sica de Sistemas
Complejos (BIFI). We have taken our measurements after every $40$ EMCS. The
total simulation time for this set of lattices has been equivalent of $0.2$
years of a Pentium IV processor running at $3.2$ GHz.  Our main effort in
$d\!=\!3$ has concerned the large, $L\!=\!16$ lattice and has been simulated
in a single FPGA (see~\ref{FPGA} for details). The total simulation time
corresponds to almost $22$ years of Pentium IV at $3.2$ GHz. Table
\ref{tab:sim_3d} shows the details of the computation.
\begin{table}[b]
% \begin{center}
\begin{ruledtabular}
\begin{tabular}{ccccccc}
 $L$ & $\beta$ & $N_\mathrm{samples}\times 10^2$ & EMCS$\times
 10^6$& EMCS/meas. \\
 \hline
 \hline
 8 & 1.41 & 10 & 4 & 40 \\
 \hline
 8 & 1.44 & 10 & 4 & 40 \\
 \hline
 8 & 1.5 & 10 & 4 & 40 \\
 \hline
 12 & 1.41 & 20 & 6 & 40 \\
 \hline
 16 & 1.385 & 2.8 & 60 & $5\times 10^5$ \\
 \hline
 16 & 1.395 & 8.5 & 60 & $5\times 10^5$ \\
 \hline
 16 & 1.405 & 10 & 60 & $5\times 10^5$ \\
 \hline
 16 & 1.41 & 2.5 & 200 & $5\times 10^5$ \\
 \hline
 16 & 1.44 & 4.8 & 500 & $5\times 10^5$ \\
 \hline
 16 & 1.5 & 0.5 & 1000 & $10^6$ \\
\end{tabular}
 %\end{center}
\end{ruledtabular}
\caption{Same as Table~\ref{tab:sim_3d} for $d\!=\!4$.}\label{tab:sim_4d}
\end{table}

In the $d\!=\!4$ model, lattices with $L\!=\!8$ and $L\!=\!12$ have been
simulated at the BIFI Cluster. The total simulation time has been the
equivalent to about $3$ years of Pentium IV at $3.2$ GHz. Again, the core of
the simulation corresponds to lattice $L\!=\!16$, and has been computed with
the FPGA. The total simulation time has been about $300$ years-equivalent of
Pentium IV. Measurements have been made every $5\times 10^5$ EMCS. The details
of the computation are shown in Table~\ref{tab:sim_4d}.

%%%%%%%%%%%%%%%%%%%%%%%%%%%%%%%%%%%%%%%%%%%%%%%%%%%%%%%%%%%%%%%%%%
\subsection{Thermalization tests}\label{THERMALIZATION}

This large computer effort has allowed us to thermalize in the broken
phase lattices of volume including up to $65536$ spins (a large
number). The thermalization issue is crucial in spin-glasses, and we
have checked it by several independent tests.

\begin{figure}
  \centerline{
\includegraphics[angle=270,width=\columnwidth,trim=10 20 10 40]{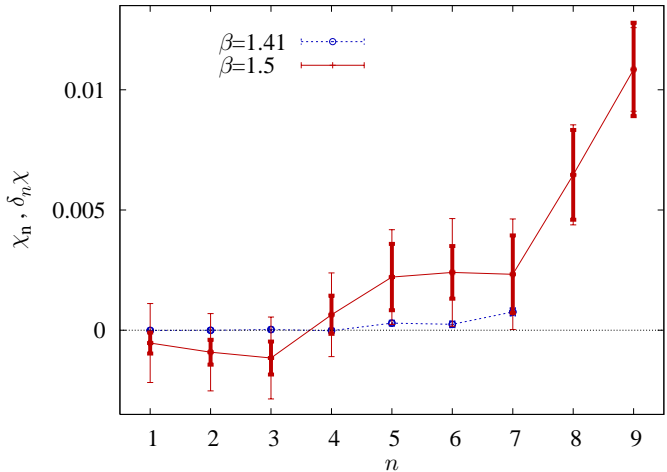}
}
\caption{(Color online). Logarithmic data binning analysis (see text) of the
  non-connected susceptibility for the $d\!=\!4$ model, $L\!=\!16$,
  $\beta\!=\!1.41$ and $\beta\!=\!1.5$. Notice that the large time
  region appears on the left in the figure.}
\label{fig:log2datab}
\end{figure}

As a first tool we have used a logarithmic binning procedure. Let us
say that during a Monte Carlo simulation we have collected estimates
for an observable quantity $O$ at all integer times $t$ in the
interval $[0,T)$. We divide these data in \textit{bins}
$I_n\!=\![T/2^{n+1},T/2^{n})$ for $n\!=\!0,1,2,3,\ldots$.  The usual
disorder average of $O$, $\overline{\langle O \rangle}$, is obtained
(after assuming that all data are at equilibrium) by averaging all
Monte Carlo data, i.e. the data over all bins. Information about
thermalization can be obtained by averaging separately over samples
the time series in the different bins. We get in this way the
logarithmic running disorder averages $O_n\equiv\overline{\langle O
  \rangle}_n$. In usual logarithmic data binning, if thermalization
has been achieved, one expects that $O_n$ becomes $n$-independent for
small $n$ (the last bins).  We show this quantity (shifted by $O_0$
for a better comparison with $\delta_n O$, see below) in the case of
the non-connected susceptibility as a function of the logarithmic
binning level $n$ in Figure \ref{fig:log2datab}. The data correspond
to the four dimensional system of volume $V=16^4$, at two values of
the temperature, one very close to the critical point and one in the
low temperature phase: the errors are drawn with a thin line.

An even better control of the convergence with time to the asymptotic
result can be obtained by computing the difference of the thermal
expectation value in bin $n$ and the value in bin $0$ in each sample,
and averaging this quantity over the disorder. In other words we
define $\delta_nO\!\equiv\!\overline{\langle O \rangle_n-\langle O
  \rangle_0}$. This way, one can obtain much smaller statistical
uncertainty: we plot this quantity for the non-connected
susceptibility in Figure \ref{fig:log2datab} by drawing the errors
with thick lines.

For both $\beta$ values of Figure \ref{fig:log2datab} both indicators
show that convergence has been reached. Errors in $\delta_n\chi$
(thick error bars) are much smaller, but they still show that the last
part of our samples has reached a steady state (even if the error is
very small all the data of the last bin are at the level of one
standard deviation from zero: also notice that the data for different
data bins are correlated, that implies that correlated discrepancies
have to be expected). We can claim that the data of the $n\!=\!0$ bin
are surely well thermalized, and we use them for computing the
equilibrium expectation values that we discuss in this note.

We have also estimated the integrated autocorrelation time $\tau$ for
the observables that we have measured: we want to be sure that the
total time length of our numerical simulation is far larger than
$\tau$.

In $d\!=\!3$, for our larger system, $L\!=\!16$, at $\beta\!=\!2.4$ (a
high value of $\beta$, deep inside the broken phase), we find that for
the internal energy $\tau\!=\!5\times10^7$ EMCS (and it turns out to
be smaller for the other observables).  This implies that our
numerical simulation has been running for a time close to $12\tau$.
In $d\!=\!4$, the length of the numerical simulation of the $L\!=\!16$
system at $\beta$ values close to the critical point turns out to be
close to $10\tau$.

\begin{figure}
\centerline{
\includegraphics[angle=270,width=\columnwidth,trim=10 20 10 42]{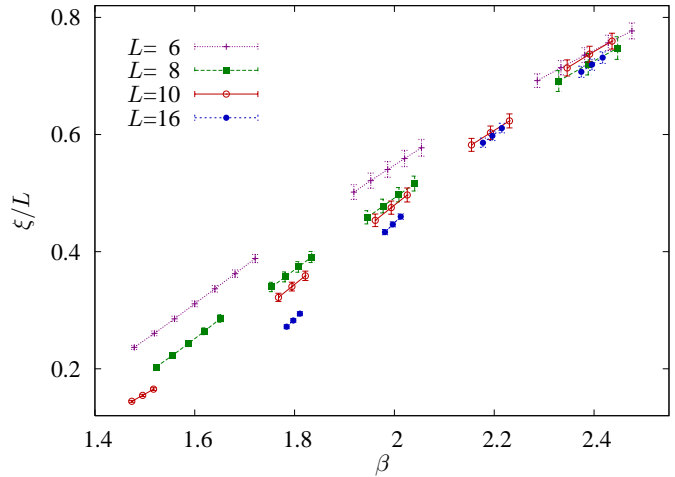}}
\caption{(Color online). Correlation length in units of the linear
size $L$ as a function of $\beta$ for $d\!=\!3$ systems of different
volumes.}
\label{fig:xi_L_3d}
\end{figure}

\begin{figure}
\centerline{
\includegraphics[angle=270,width=\columnwidth,trim=10 20 10 42]{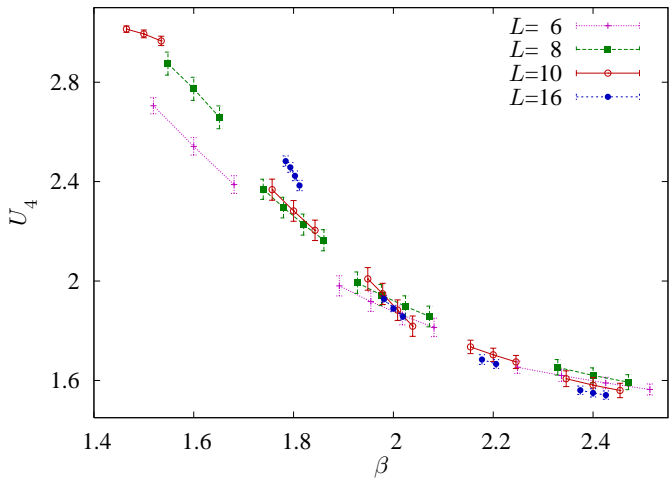}}
\caption{(Color online). The 
cumulant $U_4$ as defined in Eq.~\ref{def_binder}
as a function of $\beta$ for $d\!=\!3$ systems of different
volumes.}
\label{fig:binder_3d}
\end{figure}

We have also used a further test of thermalization, by considering the
data of the $n\!=\!0$ bin. We have done that by selecting a set of $\beta$
values to use as starting points of the reweighting extrapolation.
\cite{BALLESTEROS} Figures \ref{fig:xi_L_3d}
and \ref{fig:binder_3d} show an example of how data originated from
different disorder samples and independent numerical simulations yield
consistent results.  The choice of using different set of samples for
different $\beta$ values (the starting points of the different
reweightings that appear in the figure as neighboring groups of
points of the same type) does not optimize the quality of the final
extrapolation of the data (in the full $\beta$ interval that we consider),
but gives a further check of both the quality of the thermalization
and of the quality of the sample average. In our case the test is
obviously successful.

\begin{figure}
  \centerline{
\includegraphics[angle=270,width=\columnwidth,trim=10 20 10 42]{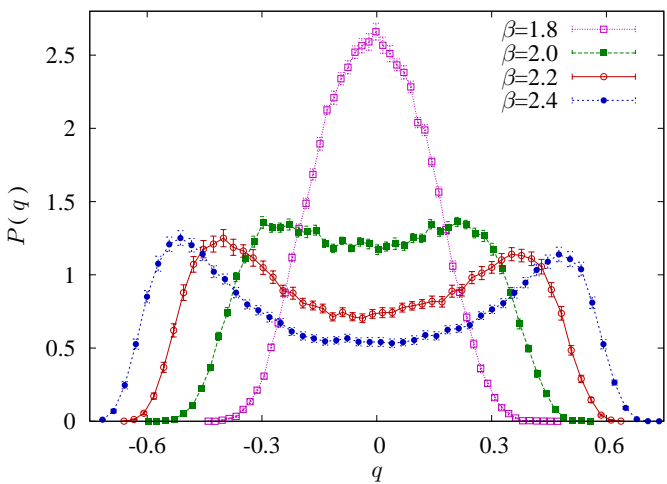}
}
\caption{(Color online). Distribution of the overlap 
in the $d\!=\!3$,  $L\!=\!16$ system, at several temperatures.}
\label{fig:qfrec_3d}
\end{figure}

\begin{figure}
  \centerline{ \includegraphics[angle=270,width=\columnwidth,trim=10 20 10
    42]{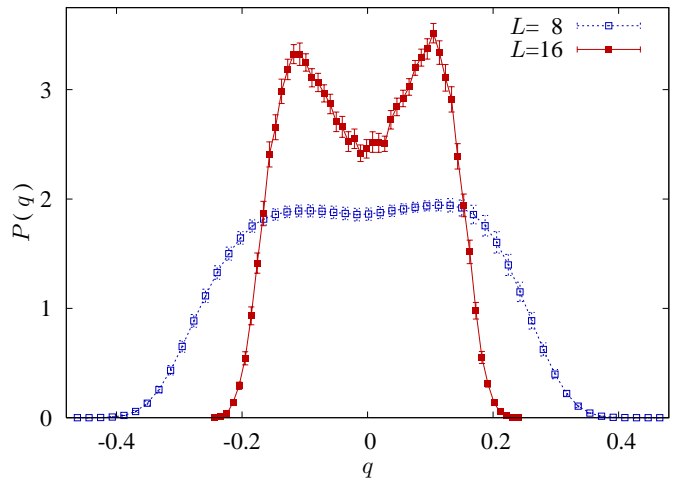} }
\caption{(Color online). Distribution of the overlap 
in the $d\!=\!4$ model at low temperature ($\beta\!=\!1.44$) for
two different lattice sizes.}
\label{fig:qfrec_4d}
\end{figure}

Even if these general thermalization checks are very useful, and they
give strong hints that the system is thermalized, the $Z_2$ symmetry
of the model (see section~\ref{MODEL}), that has been introduced
exactly with this goal in mind, is crucial to check thermalization.
Let us repeat that the allowed couplings have been selected exactly such
that the probability distribution of the modified overlap,
$P(q)$, has to be symmetric at equilibrium. We show in
Figs.~\ref{fig:qfrec_3d} and \ref{fig:qfrec_4d} $P(q)$ for $d\!=\!3$
and $d\!=\!4$ (computed by using the data of the $n\!=\!0$ bin, i.e. the last
half of the data of the numerical simulation).  These disorder
averaged distributions show very clearly the expected symmetry.

\begin{figure}
  \centerline{
    \includegraphics[angle=270,width=\columnwidth,trim=10 20 10 42]{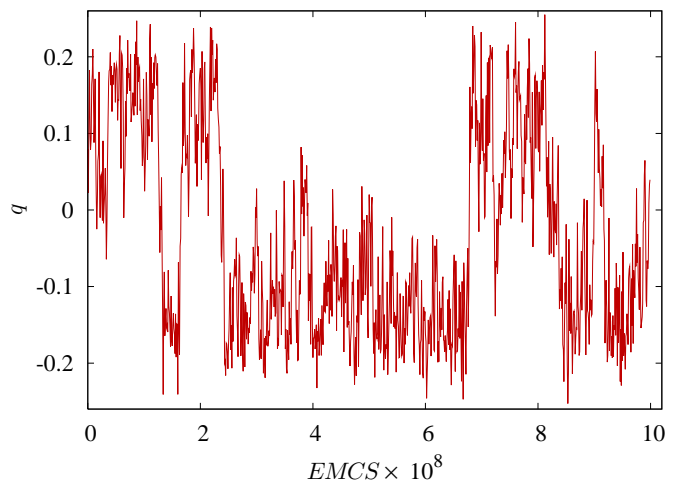}
}
\caption{(Color online). Evolution of the overlap of a representative sample
of the $d\!=\!4$ model, $L\!=\!16$ system. Here $\beta\!=\!1.5$.}
\label{fig:sample}
\end{figure}

At last we have also studied the dynamics of different observables
(for example of the modified overlap) in individual samples, and we
show an example in Fig.\ref{fig:sample}.  We can observe a number of
complete reversals of the global modified overlap, that gives us a new
estimate of the time scale on which the system gets modified: this
time scale is compatible with what we have estimated before.
We stress again that the determination of this time scale
is further evidence that we are indeed at thermal equilibrium.

We believe that this discussion clearly shows that it is
safe to use for an equilibrium analysis the data from the
$n\!=\!0$ bin (i.e. the last half of the simulation), 
since it is fully thermalized.

%%%%%%%%%%%%%%%%%%%%%%%%%%%%%%%%%%%%%%%%%%%%%%%%%%%%%%%%%%%%%%%%%%%%
\section{Results for $d\!=\!3$ model}\label{RESULTS3d}

We show in Fig.~\ref{fig:xi_L_3d} the correlation length in units of
$L$ as a function of $\beta$ for the three-dimensional model.  In the
high-temperature regime the curves for different lattice sizes are
well separated: for increasing $\beta$ the different curves approach,
and for values of $\beta$ close to $2.3$ they seem to have merged in a
single curve. In the limits of our statistical accuracy, we do not see
any sign of a splitting of the curves in the high-$T$ phase (a
crossing point at $\Tc$ and a splitting in both the low $T$ and in the
high-$T$ phase is the usual signature of a usual phase transition):
such a merging (without an eventual splitting) for increasing $\beta$
is what would happen in a Kosterlitz-Thouless transition (KT, see for
example Ref.~\onlinecite{KOSTERLITZ-THOULESS}).

The first (of the many) delicate issue about this potential behavior
concerns thermalization of the system: we have to be sure that we are
not being mislead by the fact that we have not thermalized the larger
lattice sizes (this could produce an effect hiding a crossing in the
high-$\beta$ region). This is why we have studied, and discussed
before, thermalization in detail: the thermalization checks described
in Section~\ref{MONTECARLOSECT} make us confident that we have reached
equilibrium for all the lattice sizes that we have considered. We should
not forget that there are other possible issues that could hide from
us, even in a very large scale simulation like the one discussed here,
the asymptotic result: we could need for example a better statistical
accuracy to discriminate a weak crossing, or we could need large
lattices to see the crossing appearing, or we could need to go to
higher $\beta$ values.  The issue of a very weak transition is a very
delicate one, and reliable statements must be phrased with great
care. Here we claim that a KT scenario is a possible choice given
the data that we have been able to measure in $d\!=\!3$,

In a KT scenario the quantity $\xi/L$ is expected to remain invariant
in a finite low-temperature region adjacent to the critical point. One
way to be quantitative about that is to compute the crossing points
$\beta_\mathrm{c}^{L_1,L_2}$ for the dimensionless quantity $U_4$, see
Eq.~\ref{ajuste}. In Fig.~\ref{fig:binder_3d}, we plot the cumulant
$U_4$ for several lattice sizes. The curves for different lattice
sizes cross close to $\beta\!=\!2.0$ (look for example at the
$L\!=\!8$ and the $L\!=\!16$ lattices), at a temperature where the
curves for $\xi/L$ on different lattice sizes did not yet merge (i.e.
where the correlation length has the high-$T$ behavior). The region of
the crossing is quite narrow, so that is very implausible that the
scaling corrections to $U_4$ (usually larger than that of $\xi/L$)
will shift the crossings as much as to get them close to
$\beta\!=\!2.4$. Therefore, under our numerical accuracy, we do
observe that $\xi/L$ remains invariant in an interval of temperatures
lower than that of the crossings of the cumulant.

The features we have described are consistent with a transition of the
KT type.\cite{KOSTERLITZ-THOULESS} Nevertheless, as we have discussed
before, many possible effects could lead to difficult conclusions (for
example the value of the lower critical dimension to be slightly
smaller than three). It is clear, in any case, that in $d\!=\!3$ we are
indeed sitting very close to the lower critical dimension.

%%%%%%%%%%%%%%%%%%%%%%%%%%%%%%%%%%%%%%%%%%%%%%%%%%%%%%%%%%%%%%%%%%%%%%%%
\section{Results for the  $d\!=\!4$ model}\label{RESULTS4d}

The authors of Ref.~\onlinecite{GLASSY-POTTS}, where the CRPPG model
that we investigate here was proposed, found that the four-dimensional
CRPPG undergoes a transition to a spin-glass phase at
$T\!\approx\!1.5$ (by analyzing lattices of size $L\!=\!4$ and $5$).

\begin{figure}
\centerline{ 
\includegraphics[angle=270,width=\columnwidth,trim=10 20 10 42]{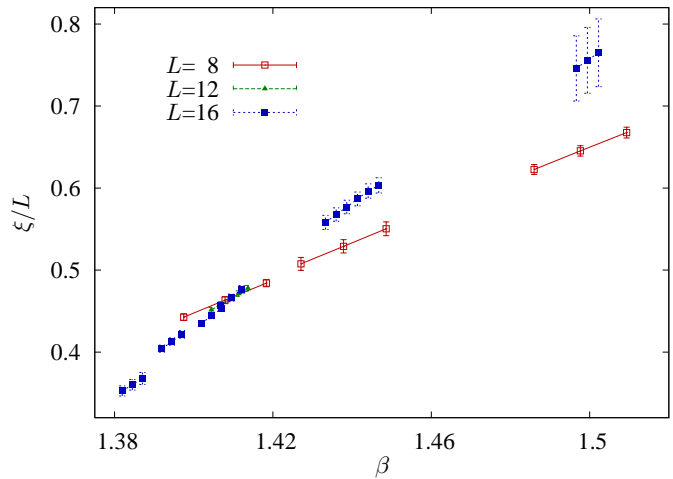} }
\caption{(Color online). Correlation length in units of $L$ as a
function of $\beta$ in the $d\!=\!4$ model.}
\label{fig:xi_L_4d}
\end{figure}

In order to analyze the transition, we study here the scaling behavior
of quantities as $\xi/L$ and $U_4$, that are expected to be
$L$-independent at the critical point.  In Fig.~\ref{fig:xi_L_4d} we
plot the correlation length in units of the lattice size as a function
of $\beta$.  The reweighting extrapolations of these quantities for
pairs of lattices $L_1$ and $L_2$ do intersect in the region around
$\beta\!=\!1.41$.  In order to be sure of the existence of the
crossing we have thermalized lattices of linear size $L\!=\!8$ and
$L\!=\!16$ deep in the low-temperature region: the normalized
correlation length of the larger lattice is well above the one of the
smaller lattice for $\beta$ values ranging from $1.44$ to $1.5$.

\begin{figure}
\centerline{
\includegraphics[angle=270,width=\columnwidth,trim=10 20 10 42]{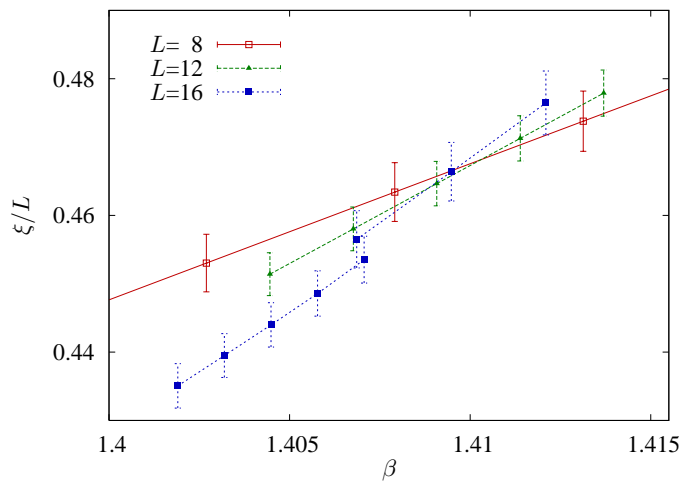}
}
\caption{(Color online). Zoom of the data of Fig.~\ref{fig:xi_L_4d}
close to the estimated critical point.}
\label{fig:xi_L_4d_cortes}
\end{figure}

In Fig.~\ref{fig:xi_L_4d_cortes} we zoom the region closer to our
putative crossing.  In this region we have also thermalized lattice of
linear size $L\!=\!12$, and we include the $L\!=\!12$ data in the figure and
in our analysis.

In Table \ref{tab:crit_4d} we give the values of the crossing points
$\beta_\mathrm{c}^{L_1,L_2}$ obtained by the crossing of the $\xi/L$
curves.  Already from Fig.~\ref{fig:xi_L_4d_cortes} it is clear that
the accuracy of the size-dependent estimates
$\beta_\mathrm{c}^{L_1,L_2}$ is not high enough to allow to estimate
scaling corrections.  This is since reaching thermal equilibrium for
$L>16$ was not in the scope of our numerical simulation (bound to run
on a single FPGA chip), while lattices with linear size $L<8$ would
have probably been too small to show true asymptotic scaling
corrections.

\begin{table}[b]
% \begin{center}
\begin{ruledtabular}
\begin{tabular}{ccccccc}
 $L_1$ & $L_2$ & $\beta_{\mathrm{c},\,\xi/L}^{L_1,L_2}$ & $\xi^*/L$ & $\nu$ & $\eta$ & $\gamma$\\
 \hline
 \hline
 8 & 12
 & 1.41(1) & 0.47(2) & 1.1(1) & -0.35(3) & 2.6(2) \\
 \hline
 8 & 16
 & 1.41(1) & 0.47(1) & 1.1(2) & -0.33(2) & 2.5(4) \\
 \hline
 12 & 16
 & 1.41(1) & 0.46(2) & 1.0(4) & -0.29(5) & 2.4(9) \\
\end{tabular}
 %\end{center}
\end{ruledtabular}
\caption{Our best estimates for the size dependent effective
critical coupling and for a number of universal quantities, as
obtained from $(L_1,L_2)$ pairs.  $\gamma$ is obtained from the
hyperscaling relation $\gamma\!=\!\nu(2-\eta)$. }\label{tab:crit_4d}
\end{table}
 
Since the cumulant $U_4$ scales like $\xi/L$ at the critical point, it
might have played the same role than $\xi/L$ (by using
Eq.~\ref{ajuste}).  However, we find that it has much larger scaling
corrections than $\xi/L$, and that these corrections shift the
crossing points to higher temperatures, out of the range that we have
analyzed (and where we believe the real asymptotic critical behavior
can be observed). We have therefore not used $U_4$ in our study of the
critical point. Our results compare fairly with the ones obtained in
Ref.~\onlinecite{GLASSY-POTTS} by analyzing systems of linear sizes
$L\!=\!4$ and $L\!=\!5$ ($\beta$ must be renormalized since our
Hamiltonian differs by a factor $2$ from the one of
Ref.~\onlinecite{GLASSY-POTTS}).

To obtain the critical exponents we consider the operators
$\partial_\beta \xi$ and $\chi$, whose associated exponents, see
Eq.~\ref{QUOTIENTS-FORMULA}, are $x_{\partial_\beta \xi}\!=\!\nu+1$
and $x_{\chi}\!=\!\gamma\!=\!\nu (2-\eta)\,$. Taking the logarithm of
the quotients of these expectation values at the crossing points of
$\xi/L$, we obtain the effective size-dependent exponents that we show
in Table~\ref{tab:crit_4d}. We can summarize our best estimate for the
$d=4$ exponents as $\beta_c=1.41(1)$, $\xi^*/L=.47(2)$, $\nu=1.1(2)$,
$\eta=-0.31(3)$ and $\gamma=2.5(4)$: these error are statistical in
nature and cannot, obviously, fully take care of the systematic
effects.

As was happening in the determination of the value of the critical
coupling, the estimated exponents lack the precision necessary for
obtaining a reliable infinite volume extrapolation.
Ref.~\onlinecite{GLASSY-POTTS} was quoting a value of $\gamma$ in the
range between $1.3$ and $1.5$, obtained from the study of the
overlap susceptibility in the warm phase of a lattice
$L\!=\!8$. Although our estimate is not very close to this value,
it is clear that we are still dealing with lattice of intermediate
size, and that a careful analysis of scaling corrections, that we hope
will soon be possible, will probably lead to reconcile these results.
Our results should characterizes, if universality holds, the spin
glass transition to a Potts Glass, independently from the detailed
model one selects. 

Finally, we also show in table~\ref{tab:crit_4d} the finite-size
estimates of the universal quantity $\xi^*/L$, i.e.  $\xi/L$ evaluated
at the critical coupling.

%%%%%%%%%%%%%%%%%%%%%%%%%%%%%%%%%%%%%%%%%%%%%%%%%%%%%%%%%%%%%%%%%%%%%%%%%
\section{Conclusions}\label{CONCLUSIONS}

We have presented a numerical study of the $4$-state CRPPG in
$d\!=\!4$, and, for the first time, in $d\!=\!3$: we have used Monte
Carlo simulations, reweighting techniques and a finite size scaling
analysis. In $d\!=\!3$ our evidence clearly shows that we are very close to
the lower critical dimension, and suggests that a Kosterlitz-Thouless
like behavior is possible, even if we could be dealing with a transient effect.
In $d\!=\!4$ we are able to collect a large number of thermalized samples 
for systems defined on large lattices, of linear size $L\!=\!16$. 
Thanks to such a large scale numerical simulation
we are able to qualify the
spin-glass transition first
found in Ref.~\onlinecite{GLASSY-POTTS}, and we
obtain size-dependent
estimates of the critical coupling, 
of the critical exponents $\nu$ and $\eta$
and of the scale-invariant quantity $\xi^*/L$

In both cases, the use of a FPGA gives us the power needed to
achieve thermalization, a target very ambitious for standard
computers.  We have been very careful in checking thermalization, and
also thanks to the built-in symmetry of the CRPPG we have succeeded in
this task.

%%%%%%%%%%%%%%%%%%%%%%%%%%%%%%%%%%%%%%%%%%%%%%%%%%%%%%%%%%%%%%%%%
\section*{Acknowledgments}

Numerical computations have been performed at BIFI. We acknowledge
partial financial support from CAM-UCM and UCM-BSCH, and from MEC
through research contracts FIS2006-08533-C03 and TEC2007-64188. J.~L.
Velasco is a DGA fellow. We thank Stefano Mossa, Giorgio Parisi and
Cristina Picus for a number of conversations about the Glassy Potts
models and more.  We thank Raffaele Tripiccione and all the JANUS
Collaboration for a continuous help that could not have been more
important for us.

%%%%%%%%%%%%%%%%%%%%%%%%%%%%%%%%%%%%%%%%%%%%%%%%%%%%%%%%%%%%%%%%%%%%%
\appendix
\section*{The FPGA device}\label{FPGA}

The problem of the glassy state, for example, is a typical problem of
very high complexity. A large (maybe infinite) number of time scales
is involved, and numerical simulations have to try to give hints about
dynamics at very long times: very large correlation and thermalization
times imply that, already on lattices of medium size, a huge
computational effort is required.  This is a typical situation where
conventional computers could be not enough to do the job.

The use of FPGA programmable chips for the simulation of spin systems
has been proposed several years ago\cite{SUE}: conventional computers
are not optimized towards the computational tasks relevant for our
typical calculation, and a FPGA can be programmed (at run time) in
order to optimize the execution of the specific problem that one wants to
solve. 

FPGA devices comes with numerous embedded and sizable memory blocks
(RAM blocks), and thousands of configurable logic blocks with programmable
interconnections. A configurable logic blocks can be programmed to
perform complex logic operations and provide storage (flip-flop
registers) at the same time.

A number of features that characterize our model are indeed optimal
for being dealt with by a FPGA: we have discrete variables that can
take a small number of values (four for our $p\!=\!4$ system), and the
interaction is local in physical space.  The Metropolis algorithm and
the random number generators discussed in Section~\ref{COMP_DETAILS}
have been implemented in the FPGA in a very effective way.

RAM blocks have a natural $2D$ (width $\times$ depth) grid
structure. A $3D$ cubic matrix of bits can be obtained by
\emph{stacking} many of them, and access to all of them with the same
memory address corresponds to addressing an entire plane in a $3D$
grid. We consider one such structure per each bit needed to represent
fields (and interactions) defined on the sites of a simple cubic
lattice.

Locality of interactions (nearest neighbors) allows for a high grade
of internal parallelism: in a checkerboard scheme, all black or all
white sites of a lattice plane can be updated simultaneously (i. e. at
the same clock cycle). Moreover, when simulating two real replicas and
mixing black (white) sites of a system with white (black) ones of its
replica, all sites in a plane can be processed in
parallel. Simultaneous local updates can then be performed by
replicating small computation cells, each executing the few simple
logical operations to compute local energies, and including a 32 bit
comparator for the Metropolis test. Precomputed transition
probabilities (that allows to
avoid lengthy computations of transcendental functions)
are stored as several small look-up tables in configurable logic
(\emph{distributed} RAM), and addressed by the computed energy
variations values (each look-up tables serves two distinct computation
cells). The iterative processes involving 32 bit integer arithmetics
for random number generators have also been \emph{parallelized}, by
cascading many 32 bit integer adders and \emph{xor}s, and allowing for
the generation of hundreds of 32 bit random numbers per clock
cycle. For further details, see Ref.~\onlinecite{IANUS2}.

We use the FPGA device Virtex 4/LX200, manufactured by Xilinx.
Depending on lattice size and number of parallel updates (between $64$
and $256$) our designs run at clock speeds between $50$ and $100$ MHz.

In Ref.~\onlinecite{IANUS2} its performances have been compared with
the ones of a $3.2$ GHz Pentium IV device: for the $d\!=\!3$ model the
FPGA performs 1800 times faster than a Pentium, while this factor is
$2300$ in $d\!=\!4$.

%%%%%%%%%%%%%%%%%%%%%%%%%%%%%%%%%%%%%%%%%%%%%%%%%%%%%%%%%%%%%%%%%%%

\end{document}